\documentclass[aps,prb,reprint,longbibliography]{revtex4-2}
\usepackage{graphicx}
\usepackage{amsmath}
\usepackage{amssymb}
\usepackage{hyperref}
\usepackage{gensymb}
\begin{document}
\title{Large in-plane negative piezoelectricity and giant nonlinear optical susceptibility in elementary ferroelectric monolayers}
\author{Ziwen Wang}
\author{Shuai Dong}
\email{sdong@seu.edu.cn}
\affiliation{Key Laboratory of Quantum Materials and Devices of Ministry of Education, School of Physics, Southeast University, Nanjing 211189, China}
\date{\today}

\begin{abstract}
Negative piezoelectrics contract in the direction of applied electric field, which are opposite to normal piezoelectrics and rare in dielectric materials. The raising of low dimensional ferroelectrics, with unconventional mechanisms of polarity, opens a fertile branch for candidates with prominent negative piezoelectricity. Here, the distorted $\alpha$-Bi monolayer, a newly-identified elementary ferroelectric  with puckered black phosphorous-like structure [J. Guo, {\it et al}. Nature \textbf{617}, 67 (2023)], is computationally studied, which manifests a large negative in-plane piezoelectricity (with $d_{33}\sim-26$ pC/N). Its negative piezoelectricity originates from its unique buckling ferroelectric mechanism, namely the inter-column sliding. Consequently, a moderate tensile strain can significantly reduce its ferroelectric switching energy barrier, while the compressive strain can significantly enhance its prominent nonlinear optical response. The physical mechanism of in-plane negative piezoelectricity also applies to other elementary ferroeletric monolayers.
\end{abstract}
\maketitle

\section{Introduction}
Piezoelectrics, which allow the interconversion between electric signal and mechanical force, are highly interesting in physical mechanisms \cite{Fu2000,Ahart2008,Wang2006,Guo2004} and essential for microelectromechanical applications such as sonars, actuators, and pressure sensors \cite{Uchino1996,Scott2007}. Generally, piezoelectricity is characterized by the piezoelectric coefficients $e_{ij}$ and $d_{ij}$, which denote the changes of polarization in response to the lattice deformation (strain $\eta$) and applied force (stress $\sigma$), respectively.

\begin{figure}
\includegraphics[width=0.48\textwidth]{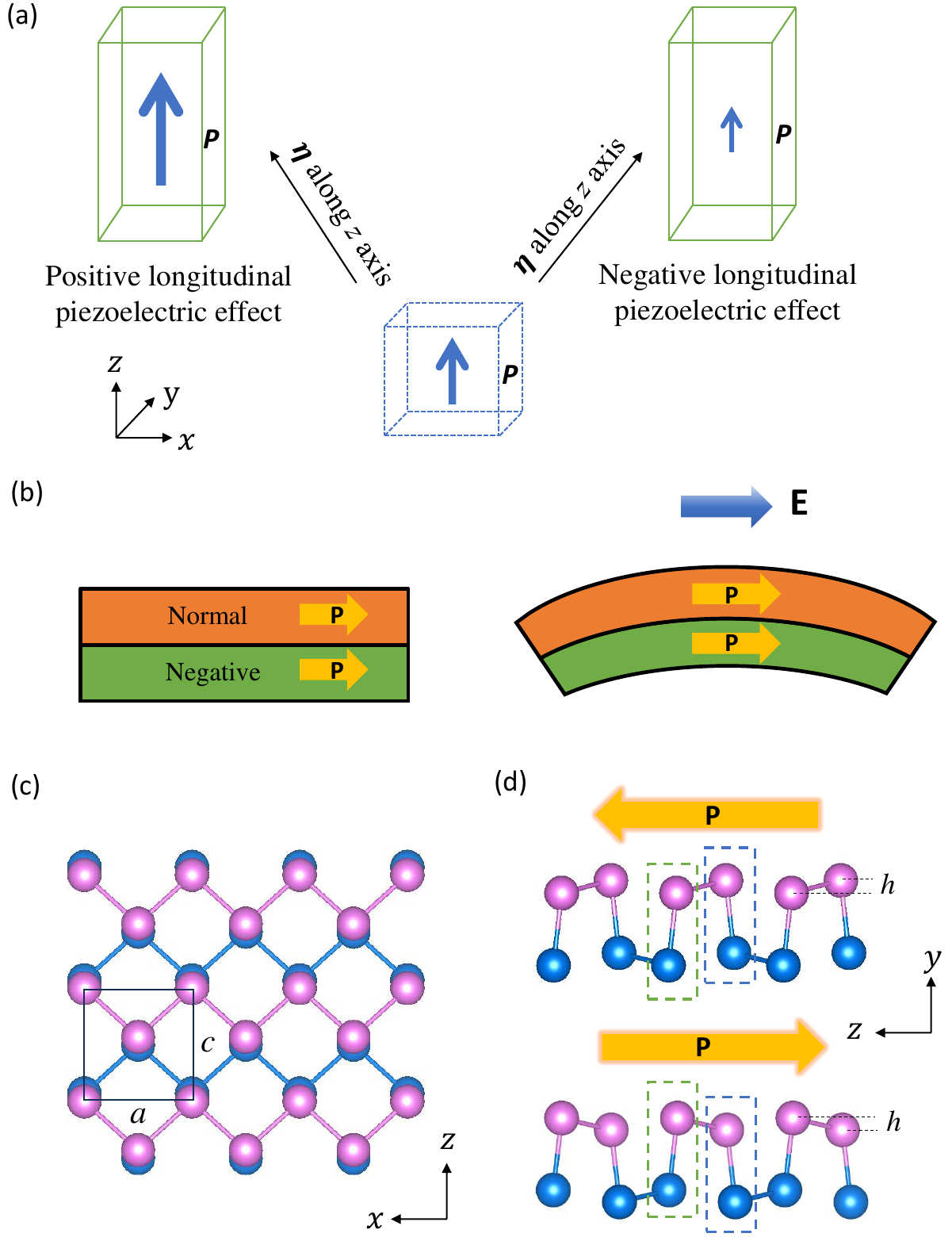}
\caption{(a) Schematic of positive and negative longitudinal piezoelectirc effects. The assumed applied strain is along the $z$-axis. (b) Schematic of the mechanical bending in a normal/negative piezoelectric heterostructure. (c-d) Structures of ferroelectric $\alpha$-Bi monolayer. The upper and lower Bi atoms are distinguished by colors. (c) Top view. The primitive cell is indicated by the black rectangle. (d) Side views of the two degenerate polar states. The buckling $h$ corresponds to the inter-column (dashed boxes) sliding. Such a sliding can induce an in-plane polarization, similar to the sliding ferroelectric mechanism in vdW layered structures. For piezoelectric tensor analysis, the in-plane polar axis is defined as the $z$-axis.}
\label{F1}
\end{figure}

Normal piezoelectrics have positive longitudinal piezoelectric coefficients ($e_{33}>0$ or $d_{33}>0$ assuming the polar axis is along $z$), namely the magnitude of polarization is more likely to increase (decrease) when a tensile (compressive) strain/stress is applied along the polar direction \cite{Duerloo2012,Fei2015}, as depicted in Fig.~\ref{F1}(a). Negative piezoelectrics with $e_{33}<0$ or $d_{33}<0$ are exotic and valuable in electromechanical system devices. For instance, by designing a heterostructure combining ultra-thin normal piezoelectric and negative piezoelectric layers, a strong bending function can be achieved, as shown in Fig.~\ref{F1}(b). However, previously only ferroelectric polymer poly(vinylidene fluoride) (PVDF) and its copolymers are few examples in this category \cite{Katsouras2016}.

Recently, a few more negative piezoelectrics were theoretical predicted or experimentally found \cite{Liu2017,Liu2020,You2019,Kim2019,Qi2021,Ding2021,Lin2019,Dutta2021}. For examples, Liu \textit{et al.} predicted several hexagonal $ABC$ ferroelectrics with negative piezoelectricity, which derive from the domination of negative clamped-ion term over the positive but small internal-strain contribution \cite{Liu2017}. You \textit{et al.} observed the out-of-plane negative piezoelectric response in ferroelectric CuInP$_2$S$_6$ and ascribed it to the reduced dimensionality of van der Waals (vdW) layered structure \cite{You2019}. Ding \textit{et al.} predicted an additional contribution to negative piezoelectricity in ZrI$_2$ vdW bulk, namely the interlayer sliding ferroelectricity is the dominated source \cite{Ding2021}. These efforts have greatly pushed forward the physical understanding of negative piezoelectricity and enlarged the scope of candidate materials. In particular, the emergence of two dimensional (2D) ferroelectrics provide a promising platform to explore negative piezoelectricity, due to their unique origins of polarity.

A latest progress of 2D ferroelectrics is the discovery of elementary ferroelectrics, which are conceptually different from traditional ferroelectric compounds invovling at least two ions (anion plus cation). In 2018, Lu {\it et al.} predicted the $\alpha$ phase As, Sb, and Bi monolayers with the puckered black phosphorous-like structures to be 2D ferroelectrics \cite{Xiao2018}, and  very recently the $\alpha$-Bi monolayer was experimentally confirmed \cite{Gou2023}. Such an exciting  branch provides great opportunities to explore the exotic dielectric properties. 

In this Letter, the piezoelectricity of elementary ferroelectric monolayers have been studied using density functional theory (DFT) calculations. Taking $\alpha$-Bi monolayer as the representive, our calculation reveals a large negative piezoelectric coefficient, which originates from its unique ferroelectric mechanism. The puckered structure can mimic the inter-column sliding, leading to a similar but much stronger effect, comparing with that in interlayer sliding ferroelectrics. Furthermore, its nonlinear optical response is found to be rather prominent, which can be further enhanced by the compressive strain.

\section{Computational methods}
DFT calculations are performed using Vienna \textit{ab initio} Simulation Pack (VASP) \cite{Kresse1996}. The projector augmented wave (PAW) pseudopotentials are Bi\_d ($5d^{10}6s^26p^3$), As ($4s^24p^3$), Sb ($5s^25p^3$), Ge\_d ($3d^{10}4s^24p^2$), Se ($4s^24p^4$), Zr\_sv ($4s^24p^65d^25s^2$), and I ($5s^25p^5$), as recommended by VASP. Plane-wave cutoff energy is fixed as $400$ eV. The default exchange-correlation functional is treated using Perdew-Burke-Ernzerhof (PBE) parametrization of the generalized gradient approximation (GGA) \cite{Perdew1996}, unless specifically stated. Other functionals have also been tested, including Perdew-Burke-Ernzerhof-revised (PBEsol) parametrization of GGA \cite{Perdew2008} and Perdew-Zunger parametrization of the local density approximation (LDA) \cite{Perdew1981,Ceperley1980}. 

The coordinates of $\alpha$-Bi monolayer is shown in Fig.~\ref{F1}(c), with the $y$-axis as the out-of-plane direction. To simulate a monolayer, a $25$ \AA{} vacuum layer is added to avoid the interaction between two neighboring slices. For Brillouin zone sampling, $\Gamma$-centered $11\times1\times11$ Monkhorst-Pack $k$-mesh are adopted for Sb, Bi and GeSe monolayers, and $11\times6\times3$ for ZrI$_2$ bulk. Both the lattice constants and atomic positions are fully optimized iteratively until the Hellmann-Feynman force on each atom and the total energy are converged to $0.01$ eV/\AA{} and $10^{-6}$ eV, respectively. 

The ferroelectric polarization is calculated using the Berry phase method \cite{King-Smith1993} and the possible ferroelectric switching path is evaluated by the linear interpolation between the optimized ferroelectric state (FE) and optimized paraelectric (PE) state. To calculate the piezoelectric stress coefficients $e_{ij}$, the density functional perturbation theory (DFPT) is employed \cite{abinit3}, with more dense $15\times1\times15$ and $13\times7\times4$ $k$-meshes sampling for monolayers and bulk, respectively. The elastic stiffness tensor matrix elements are calculated by vaspkit~\cite{wang:cpc}.

Second harmonic generation (SHG) susceptibilities are calculated by ABINIT package \cite{abinit1,abinit2,abinit3,abinit4}. A dense $k$-point sampling of 50$\times$1$\times$50, and 40 electronic bands are used for achieving the SHG susceptibility tensor.

To simulate the uniaxial in-plane strain, the lattice constant along the strain direction is fixed, while all atomic positions and other  directional lattice constant(s) are fully relaxed.  Then the changes of polarization in response to an applied  strain $\eta$ and stress $\sigma$ can be gauged by the piezoelectric stress tensor $e_{ikl}$ and piezoelectric strain tensor $d_{ikl}$, respectively. The formulas can be expressed as follows:
\begin{equation}
	e_{ikl}=\left(\frac{\partial P_i}{\partial \eta_{kl}}\right)_E=-\left(\frac{\partial \sigma_{kl}}{\partial E_i}\right)_\eta
	\label{eq1}
\end{equation}
\begin{equation}
	d_{ikl}=\left(\frac{\partial P_i}{\partial \sigma_{kl}}\right)_E=\left(\frac{\partial \eta_j}{\partial E_{kl}}\right)_\sigma
	\label{eq2}
\end{equation}
where $i$, $k$, $l$ $\in$ $\left\{1,2,3\right\}$, with $1$, $2$, $3$ corresponding to $x$, $y$, $z$. $P$ and $E$ denote the polarization and electric field, respectively. Using the Voigt notation, $e_{ikl}$ and $d_{ikl}$ can be reduced to $e_{ij}$ and $d_{ij}$ respectively, where $j$ $\in$ $\left\{1,2,3,\cdots,6\right\}$ and 1$\mapsto$ 11 ($xx$), 2$\mapsto$ 22 ($yy$), 3$\mapsto$ 33 ($zz$), 4$\mapsto$ 23 or 32 ($yz$ or $zy$), 5$\mapsto$ 13 or 31 ($xz$ or $zx$), 6$\mapsto$ 12 or 21 ($xy$ or $yx$).

\section{Results and discussion}
\subsection{Negative piezoelectricity}
The distorted $\alpha$-Bi monolayer with black phosphorous-like structure  owns an orthorhombic lattice (space group $Pmn2_1$, no. 31), as shown in Fig.~\ref{F1}(c-d). A unit cell consists of four Bi atoms, forming the upper and lower sheets. Our DFT optimized lattice constants agree well with the experimental values, as compared in Table~\ref{Tab-1}, which indicates the reliability of our calculation. 

\begin{table*}
\caption{DFT calculated basic physical properties of $\alpha$-Sb and $\alpha$-Bi monolayer, in comparison with ZrI$_2$ bulk and GeSe monolayer. The polarizations are in units of pC/m and $\mu$C/cm$^2$ for monolayers and bulk, respectively. The piezoelectric stress coefficients ($e_{ij}$) are in units of 10$^{-10}$ C/m and C/m$^2$ for the monolayers and bulk, respectively. Here the space group for all four materials is $Pmn2_1$ (no. 31). For all monolayers, the out-of-plane direction is along the $b$-axis, while for ZrI$_2$ bulk the $c$-axis is the vdW stacking direction. It should be noted that in our GGA-PBE calculation, $\alpha$-As monolayer exhibits a symmetric structure with space group $Pmna$ (no. 53), different from previous LDA result \cite{Xiao2018}. More details and tested results with different exchange-correlation functionals can be found in SM~\cite{sm}.} 
\begin{tabular*}{\textwidth}{@{\extracolsep{\fill}}lllllllll}
\hline \hline
Structure& $a$ (\AA) & $c$ (\AA) & $b$ (\AA)& Polarization & $e_{33}$ & $e_{31}$& $d_{33}$ (pC/N) &   Gap (eV)               \\
\hline
    Sb  & $4.36$  & $4.73$   & - &  $21$  & $-2.7$  & $-1.1$   & $-19.2$  &     $0.23$        \\
	Bi  & $4.57$  & $4.83$  & - &  $16$  & $-5.1$  & $-2.1$   & $-25.9$  &     $0.31$        \\
    Bi (Exp. \cite{Sun2012})	& $4.54 $   & $4.75$   & - & &       &     &   & \\
    Bi (Exp. \cite{M2019})	& $4.5$    & $4.8$   & - & &       &     &   & \\
    Bi (Cal.  \cite{Xiao2018})	& $4.39$    & $4.57$   & - & &       &     &   & \\
		ZrI$_2$   & $3.75$  & $14.81$  & $6.86$   &  $0.37$  & $-0.061$ &$-0.002$ & $-1.416$& $0.19$\\
	ZrI$_2$  (Cal. \cite{Ding2021})	& $3.75$  & $14.80$  & $6.87$   & $0.39$   & $-0.061$  & $-0.001 $  & $-1.445 $  & $0.15 $  \\
		 GeSe &$3.97$    & $4.28$ &   -  &  $360$ &  $11.5$   & $-3.3$ &  $100.1$  &    $1.27$  \\
	GeSe (Cal. \cite{Gomes2015,Fei2016})	&$3.99$ & $4.26$ & - & $367$ & $13.3$ & $-3.0$   &    & \\
\hline \hline
\end{tabular*}
\label{Tab-1}
\end{table*}

The symmetry of $Pmn2_1$ space group (point group $mm2$) allows five independent elements of piezoelectric tensor matrix: $e_{31}$, $e_{32}$, $e_{33}$, $e_{24}$, and $e_{15}$ \cite{De2015,sm}. For 2D materials, usually only the in-plane stresses and strains are allowed, while the out-of-plane direction (i.e. the $y$ axis here) is stress/strain free \cite{Dong2017,sm}, i.e., $\sigma_2=\sigma_4=\sigma_6=0$. Thus, the piezoelectric tensor matrix can be reduced as follows:
\begin{equation}
e=\begin{pmatrix}
	0  &0  &e_{15}  \\
	0  &0  & 0      \\
	e_{31}& e_{33}&0
\end{pmatrix}.
\end{equation}

Similarly, the independent elements of the elastic stiffness tensor ($C$) for 2D rectangular lattice are four ($C_{11}$, $C_{13}$, $C_{33}$, $C_{55}$) \cite{Mazdziarz2019,sm}:

\begin{equation}
C=\begin{pmatrix}
		C_{11}&C_{13} &0 \\
		C_{31}&C_{33}&0\\
	      0 &0   &C_{55}    \\ 
	\end{pmatrix}.
\end{equation}

Then the piezoelectric strain coefficients $d_{ij}$ can be calculated based on the piezoelectric stress tensor $e$ and the elastic stiffness tensor $C$, as follows:
\begin{equation}
	d_{ij}=\sum_{k=1}^{3}e_{ik}C_{kj}^{-1}.
	\label{eq3}
\end{equation}

The calculated longitudinal piezoelectric stress and strain coefficients ($e_{33}$ and $d_{33}$) of $\alpha$-Sb and $\alpha$-Bi monolayer are also summarized in Table~\ref{Tab-1}, in comparison with ZrI$_2$ bulk (an interlayer sliding vdW ferroelectric) and GeSe monolayer (a conventional ion-displacive type ferroelectric). The complete piezoelectric tensors and elastic stiffness tensors can be found in Supplemental Material (SM) \cite{sm}. Our calculated results find larger negative piezoelectric coefficients in elementary ferroelectric monolayers than that of ZrI$_2$. Furthermore, although the structure of GeSe monolayer is similar to $\alpha$-Bi monolayer, it exhibits a normal piezoelectric response, due to its different ferroelectric mechanism (to be discussed later). 

\begin{figure}
\includegraphics[width=0.49\textwidth]{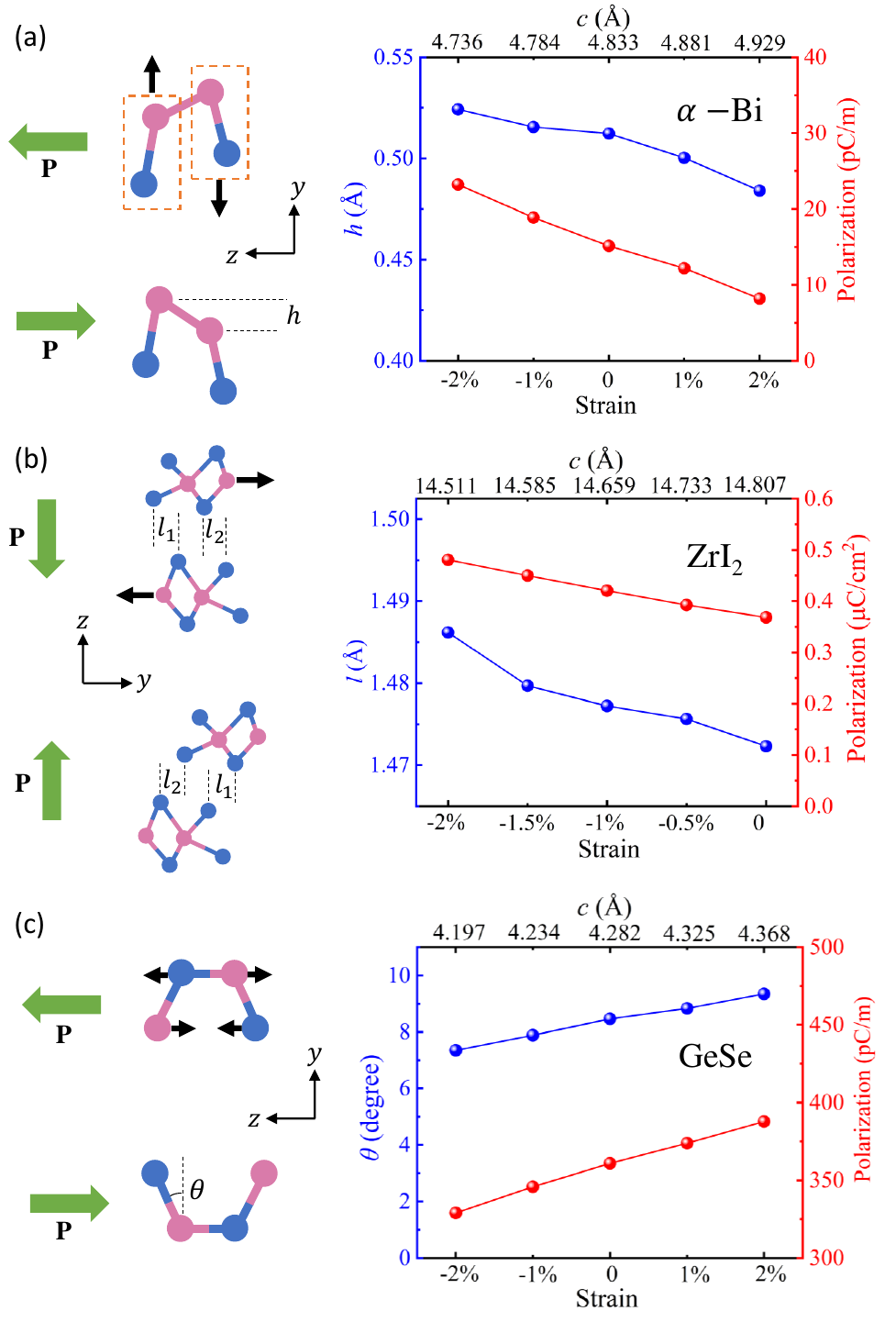}
\caption{Comparison of piezoelectricity with different mechanisms. Left: the microscopic order parameters for ferroelectric distorions: $h$, $l=l_1+l_2$, and $\theta$. Here \textit{l} is the order parameter of the sliding ferroelectric, which denotes the sliding length between the two degenerate ferroelectric states. The atomic movements during the ferroelectric switching process are indicated by black arrows. Green arrows: the vectors of polarizations. Right: uniaxial strain dependence of ferroelectric order parameters and corresponding dipole moments. The uniaxial strains are applied along their polar directions. (a) Inter-column sliding ferroelectric $\alpha$-Bi monolayer. (b) Interlayer sliding ferroelectric ZrI$_2$ bulk. (c) Ion-displacive type ferroelectric GeSe monolayer. $\alpha$-Bi monolayer and ZrI$_2$  bulk show similar evoluations of order parameter and dipole moment, for their similar sliding mechanisms, which are opposite to the GeSe monolayer.  }
\label{F2}
\end{figure}

The origin of in-plane negative piezoelectricity in $\alpha$-Bi monolayer can be traced back to its ferroelectric mechanism. As previous studies revealed \cite{Gou2023,Xiao2018}, the distorted structure with buckling $h$ breaks the centrosymmetry of $\alpha$-Bi monolayer and induces charge transfer in each puckered sheet (see orbital-projected charge density distributions of FE and PE phases in Fig.~S2 of SM~\cite{sm}), giving rise to the in-plane polarization. Such buckling makes its origin of ferroelectricity similar to recently reported sliding ferroelectricity in layered structures \cite{Li2017,Ding2021,Miao2022,Yasuda2021,Wu2021,Wang2021}, as compared in the left sides of Figs.~\ref{F2}(a-b). Taking interlayer sliding ferroelectric ZrI$_2$ as an example [Fig.~\ref{F2}(b)], the interlayer interaction in the polar stacking mode causes charge redistribution within each ZrI$_2$ layer, leading to an electric dipole along the stacking direction. By analogy, the unique buckling can be considered as a kind of inter-column sliding in $\alpha$-Bi monolayer. Keeping this similarity in mind, the negative piezoelectricity of $\alpha$-Bi monolayer is natural. Namely, with compressive strain along the polar axis,  the neighbor columns become closer. Such squeeze effect strengths the polarization originating from inter-column sliding, as appears in interlayer sliding ferroelectric ZrI$_2$ bulk \cite{Ding2021}. Our DFT calculation confirms such an exotic evolution tendency in both $\alpha$-Bi monolayer and ZrI$_2$ bulk, as shown in Fig.~\ref{F2}: the shorter lattice constant along the polar axis, the stronger ferroelectric distortion, and the larger polarization.

Despite the similarity, that interaction between the inter-column sliding which involves covalent bonds, should be much stronger than that between interlayer sliding which involves the vdW interaction. As a result, the induced negative piezoelectricity in $\alpha$-Bi monolayer is $18.3$ times of that in ZrI$_2$ bulk,  as compared in Table~\ref{Tab-1}, which is in the same order of magnitude with the so-called gaint negative piezoelectricity of CuInP$_2$S$_6$ (experiment $\sim$ $-95$ pC/N and DFT $\sim -18$ pC/N) \cite{You2019}. Note that the gaint negative piezoelectricity of CuInP$_2$S$_6$ is out-of-plane due to the soft vdW layer, while that of our $\alpha$-Bi monolayer is in-plane, with totally different mechanisms.

For GeSe monolayer with conventional ferroelectric origin, though its structure is similar to the $\alpha$-Bi monolayer, the compressive (tensile) strain along the polar axis can only reduce (increase) its microscopic ferroelectric order parameter $\theta$, and thus generally exhibits a normal piezoelectricity, as shown in Fig.~\ref{F2}(c).

\begin{table}
	\caption{Calculated clamped-ion ($\bar{e}_{ij}$) and internal-strain (${e}_{ij}'$) piezoelectric stress coefficients of Bi and GeSe monolayers. The piezoelectric coefficients are in units of $10^{-10}$ C/m. }
	\begin{tabular*}{0.48\textwidth}{@{\extracolsep{\fill}}lcccccc}
		\hline \hline
		& \multicolumn{3}{c}{Clamped-ion} & \multicolumn{3}{c}{Internal-strain}             \\ \cline{2-4}  \cline{5-7} 
		& $\bar{e}_{31}$ &$\bar{e}_{33}$ &$\bar{e}_{15}$ & ${e}_{31}'$ & ${e}_{33}'$ & ${e}_{15}'$\\ 
		\hline
		Bi       & $2.6$   & $0.6$  & $1.3$ &  $-4.7$ & $-5.7$&  $-5.6$ \\ 
		GeSe     &$-7.1$    & $-0.3$  & $-7.1$ &  $3.9$  &$11.9$&$14.0$ \\
		\hline \hline
	\end{tabular*}
	\label{Tab-2}
\end{table}

Following the analysis of Refs.~\cite{PRL1998,Dutta2021}, the piezoelectric stress coefficients $e_{33}$ can be decomposed into two parts: the clamped-ion term $\bar{e}_{33}$ and the internal-strain term $e_{33}'$. $\bar{e}_{33}$ denotes the change of polarization $P$ due to the uniform distortion of the lattice with the atomic fractional coordinates fixed, and $e_{33}'$ denotes the piezoelectric response to the atomic relaxations that release the internal strain. The decomposition of $\alpha$-Bi monolayer and GeSe monolayer calculated are summarized in Table ~\ref{Tab-2}. It is clear that the $\alpha$-Bi monolayer has a dominant negative internal-strain $e_{33}'$ contribution (detailed Born effective charge and internal strain tensors can be found in SM~\cite{sm}) whereas the clamped-ion term $\bar{e}_{33}$ is positive but small. This result is opposite to the negative piezoelectricity in the so-called $ABC$ ferroelectrics with strong ionic bonds \cite{Liu2017}, where the clamped-ion term $\bar{e}_{33}$ is negative and dominates the positive internal-strain $e_{33}'$, but similar to the vdW layered compounds BiTe$X$ \cite{Kim2019}.  In contrast, the GeSe monolayer possesses a negative clamped-ion term $\bar{e}_{33}$ while the much larger positive internal-strain $e_{33}'$ decides the total positive piezoelectricity. 

\begin{figure}
	\includegraphics[width=0.48\textwidth]{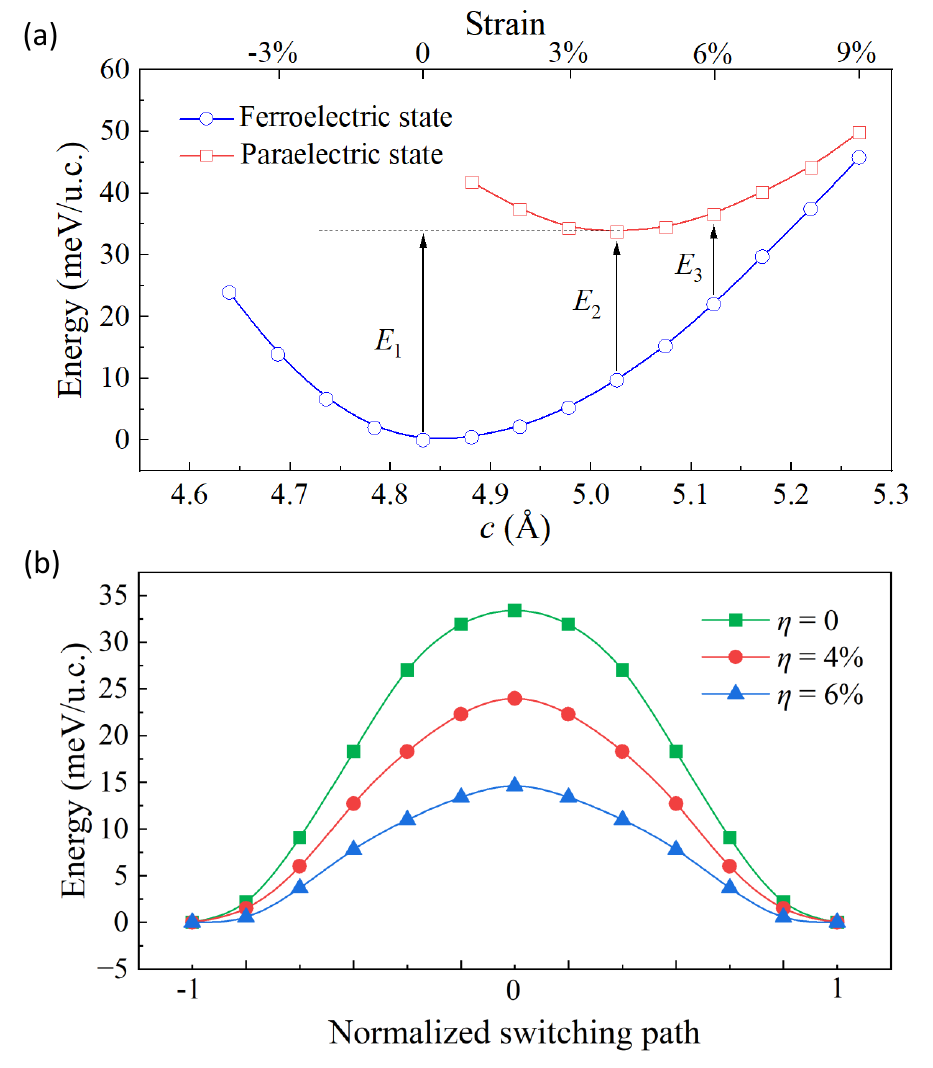}
	\caption{(a) Comparison of energy curves for the ferroelectric state and paraelectric state as a function of the lattice constant of $c$. (b) The ferroelectric switching barriers at different conditions of strains. $E_1$, $E_2$, and $E_3$ in (a) are the barrier heights in (b). }
	\label{F3}
\end{figure}

An intresting consequence of negative piezoelectricity is the contracted lattice constant along the polar axis, which was also observed in negative piezoelectric WO$_2$X$_2$ (X = Br and Cl) monolayers \cite{Lin:prl}. Figure~\ref{F3}(a) shows the energy curves of ferroelectric state and undistorted paraelectric state as a function of lattice constant $c$. As expected, the optimized lattice constant $c$ is $3.x\%$ longer in the paraelectric one, and thus a tensile strain can reduce the energy gain from ferroelectric distortion, which is beneficial to lowering energy consumption during the ferroelectric switching. As demonstrated in Fig.~\ref{F3}(b), the ferroelectric switching barriers are significantly reduced upon uniaxial tensile strains along the $c$-axis, e.g. $70\%$ lower at strain $\eta=6\%$.

The in-plane piezoelectric stress coefficients $e_{33}$ and $e_{31}$ as a function of compressive and tensile strains are also studied. In general, $e_{33}$ and $e_{31}$ increase with the uniaxial compressive strain but decrease with the tensile strain. These results can be found in Fig. S3 of SM~\cite{sm}.

\subsection{Prominent SHG signal}
The optical second harmonic generation (SHG), is a powerful tool to characterize ferroelectric materials, in particular vital for those 2D ferroelectrics while those conventional electrical methods are difficult \cite{Moqbel2022,Song2022,Ding2023,Xu2023}. In fact, although many 2D ferroelectrics have been claimed experimentally, their precise polarizations from direct electrical measurement remain challenging \cite{Guan2020}. Considering the narrow bandgap of $\alpha$-Bi monolayer ($\sim0.31$ eV in GGA calculation), the direct electrical measurement of ferroelectricity and piezoelectricity may be tough. Therefore, the optical SHG is an essential route to characterize the polarization and its change upon strain.

\begin{figure}
	\includegraphics[width=0.48\textwidth]{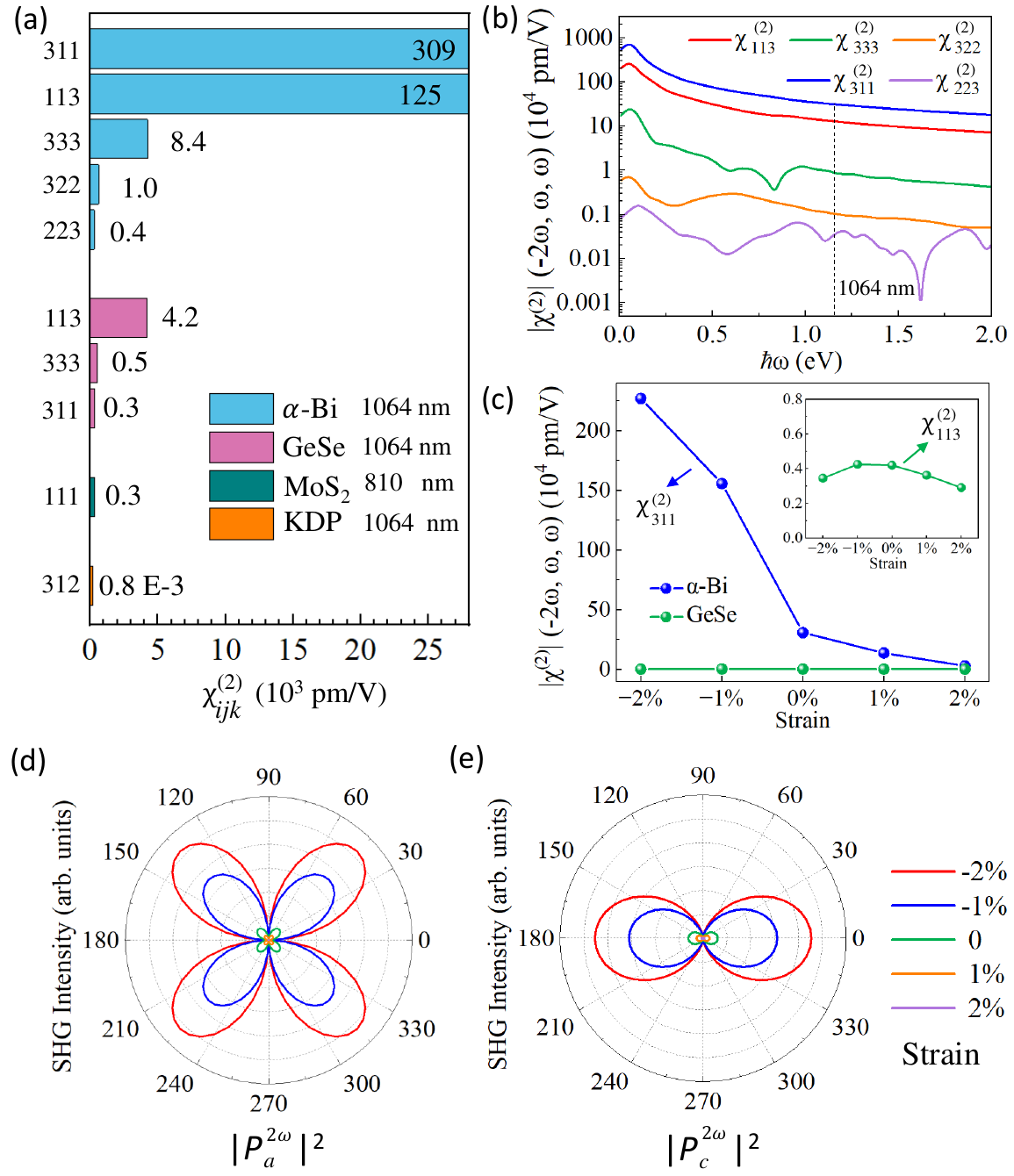}
	\caption{Calculated SHG properties of the ferroelectric $\alpha$-Bi monolayer. (a) The magnitudes of five independent SHG susceptibilities of $\alpha$-Bi monolayer, in comparison with other nonlinear optical materials. The values of MoS$_2$ at $810$ nm and KDP are taken from Refs.~\cite{Eckardt1990,Li2013}. (b) SHG susceptibilities as a function of frequency of incident light. (c) The dominant susceptibility $\chi_{311}^{(2)}$ at $1064$ nm under moderate uniaxial strains. Inset: the maximum component $\chi_{113}^{(2)}$ of GeSe monolayer at $1064$ nm for comparison. (d-e) The polar plot of SHG components under uniaxial strains, with a perpendicular incident light. The whole SHG signal is the sum of these two components.}
	\label{F4}
\end{figure}

Generally, the SHG intensity $I$ can be estimated as: 
\begin{equation}
	I \propto (P^{2\omega})^2=(P_a^{2\omega})^2 + (P_b^{2\omega})^2 + (P_c^{2\omega})^2,
\end{equation}
where $P^{2\omega}$ is the second harmonic polarization generated by the electric field $E$($\omega$) component of incident light with angular frequency $\omega$. $a$, $b$ and $c$ are the crystal orientations, as shown in Fig.~\ref{F1}(c). The components of $P_i^{2\omega}$ can be expressed as \cite{Sutherland2003}:
\begin{equation}
P_i^{2\omega}=\varepsilon_0\sum_{jk}\chi_{ijk}^{(2)}(-2\omega,\omega,\omega)E_j(\omega)E_k(\omega),
\end{equation}
where $\varepsilon_0$ is the vacuum permittivity, and $\chi_{ijk}^{(2)}$ is the SHG susceptibility. 

For $\alpha$-Bi monolayer with $Pmn2_1$ space group (point group $mm2$), there are five independent elements in SHG susceptibility tensor matrix: $\chi_{113}^{(2)}=\chi_{131}^{(2)}$, $\chi_{223}^{(2)}=\chi_{232}^{(2)}$, $\chi_{311}^{(2)}$, $\chi_{322}^{(2)}$, and $\chi_{333}^{(2)}$, while other tensor elements are rigidly zero as required by the symmetry. Using a simplified notation, $P_i^{2\omega}$ can be given by \cite{Sutherland2003}: 
\begin{eqnarray}
\nonumber \left[
\begin{array}{c}
	P_a^{2\omega}\\
	P_b^{2\omega} \\
	P_c^{2\omega}
\end{array}
\right]
&=&2\varepsilon_0
\begin{bmatrix}
	0     &      0 &    0 & 0      &d_{15} & 0\\   
	0     &      0 &    0 & d_{24} &0      & 0 \\   
	d_{31}& d_{32} &d_{33}& 0      &0      &0
\end{bmatrix}
\left[
\begin{array}{c}
	E_a^2   \\
	E_b^2   \\
	E_c^2   \\
	2E_bE_c \\
	2E_aE_c \\
	2E_aE_b 
\end{array}
\right]	\\
&=&2\epsilon_0
\begin{bmatrix}
	2d_{15}E_aE_c\\   
	2d_{24}E_bE_c \\   
	d_{31}E_a^2+d_{32}E_b^2+d_{33}E_c^2
\end{bmatrix},
\end{eqnarray}
where $d_{il}$'s are the so-called $d$-coefficient which are usually used to represent  SHG susceptibility $\chi^{(2)}$. The subscripts are linked by the intrinsic permutation symmetry, namely: $d_{il} \rightarrow d_{ijk} \rightarrow  \frac{1}{2}\chi^{(2)}_{ijk}$~\cite{Sutherland2003}.

The calculated SHG susceptibility $\chi_{ijk}$'s of $\alpha$-Bi monolayer at $\hbar\omega=1.17$ eV (i.e. wavelength $\lambda=1064$ nm which is frequently used in SHG experiments) are shown in Fig.~\ref{F4}(a), in comparison with other nonlinear optical materials. At $\hbar\omega=1.17$ eV, a giant susceptibility is obtained: $\chi_{311}^{(2)}=3.09\times10^5$ pm/V, which is much higher than GeSe monolayer ($\sim10^3$ pm/V~\cite{Wang2017,sm}),  MoS$_2$ monolayer ($\sim10^2$ pm/V at $810$ nm~\cite{sm,Wang2017,Wang2015}), and KDP ($\sim 0.76$ pm/V~\cite{Eckardt1990}, a well-known standard SHG reference). The SHG susceptibilities as a function of the light frequency are plotted in Fig.~\ref{F4}(b), which is even more larger in the low energy region.  In this sense, the distorted $\alpha$-Bi monolayer is a very promient nonlinear optical material.

The large negative piezoelectricity can be also reflected in the SHG signal. Taking the largest element $\chi_{311}^{(2)}$ as example, Fig.~\ref{F4}(c) shows its evolution under uniaxial strain. As expected, the compressive strain can enhance $\chi_{311}^{(2)}$. Surprisingly, such an enhancement is very large: almost one order of magnitude larger at $\eta=-2\%$. Therefore, the SHG can be used  as a sensitive method to monitor the strain of $\alpha$-Bi monolayer. For comparison, the strain effect to $\chi_{113}^{(2)}$ (the maximum susceptibility at $1.17$ eV) of GeSe monolayer is rather insensitive. 

With a perpendicular incident light along the $b$ axis, its electric field can be expressed as $E=(E_a, E_b, E_c)=E(\cos\varphi, 0, \sin\varphi)$. Thus, the nonzero $P_i^{2\omega}$ components can be derived as:
\begin{eqnarray}
\nonumber P_a^{2\omega} & \propto & \chi_{113}^{(2)}\sin(2\varphi),  \\
P_c^{2\omega} & \propto & \chi_{311}^{(2)}\cos^2\varphi + \chi_{333}^{(2)}\sin^2\varphi,
\end{eqnarray}
where $\varphi$ is the angle between the $E$ vector and the $a$ axis. Then the angle-dependent $P_a^{2\omega}$ and  $P_c^{2\omega}$ can be obtained under uniaxial strains along the $c$-axis, as shown in Figs.~\ref{F4}(d-e). The uniaxial strain can significantly enhance the SHG components $P_a^{2\omega}$ and  $P_c^{2\omega}$, but will not alter their four-fold symmetry and two-fold symmetry.

At last, we also examined the epitaxial strain effect to $\alpha$-Bi monolayer. The single-layer graphene was chosen as the substrate, as done in the experiment \cite{Sun2012}. After the full structural optimization, the inherent ferroelectricity of $\alpha$-Bi monolayer can be preserved by the substrate (see Fig. S5 in SM~\cite{sm}).
 
\section{Conclusion}
In summary, the elementary ferroelectric $\alpha$-Bi monolayer and its sister compounds have been systematically studied by DFT calculations, which exhibit large in-plane negative piezoelectricity ($d_{33}=-26$ pC/N) and gaint nonlinear optical susceptibility ($\chi_{311}^{(2)}=3.09\times10^5$ pm/V). Their negative piezoelectricity arises from the intriguing ``inter-column'' sliding ferroelectric mechanism, different from the negative piezoelectricity in CuInP$_2$S$_6$ and $ABC$ ferroelectrics. The prominent SHG intensity in $\alpha$-Bi monolayer can be drastically enhanced with a moderate uniaxial compressive strain, while its ferroelectric switching energy barrier can be reduced by applying an uniaxial tensile strain. Our work will encourage more theoretical and experimental works on 2D negative piezoelectricity and elementary ferroelectrics, which not only refresh the physical knowledge of polarity but also offer bright future for low dimensional electromechanical devices.

\begin{acknowledgments}
We thanks C. R. Gui and X. Y. Yang for useful discussions. Work was supported by National Natural Science Foundation of China (Grants No. 12274069 and No. 11834002) and the Big Data Computing Center of Southeast University.
\end{acknowledgments}

\bibliography{Bi}
\end{document}